\documentstyle[12pt, psfig]{article}
\renewcommand{\thepage}{\arabic{page}}
\newcommand{\nc}{\newcommand}
\nc{\beq}{\begin{equation}} \nc{\eeq}{\end{equation}}
\nc{\beqa}{\begin{eqnarray}} \nc{\eeqa}{\end{eqnarray}}
\nc{\lsim}{\begin{array}{c}\,\sim\vspace{-21pt}\\< \end{array}}
\nc{\gsim}{\begin{array}{c}\sim\vspace{-21pt}\\> \end{array}}

\newcommand{\drawsquare}[2]{\hbox{%
\rule{#2pt}{#1pt}\hskip-#2pt
\rule{#1pt}{#2pt}\hskip-#1pt
\rule[#1pt]{#1pt}{#2pt}}\rule[#1pt]{#2pt}{#2pt}\hskip-#2pt
\rule{#2pt}{#1pt}}

\newcommand{\Yfund}{\raisebox{-.5pt}{\drawsquare{6.5}{0.4}}}
\newcommand{\Yasymm}{\raisebox{-3.5pt}{\drawsquare{6.5}{0.4}}\hskip-6.9pt%
        \raisebox{3pt}{\drawsquare{6.5}{0.4}}}

\newcounter{mysection}
\newcounter{mysubsection}
\newcommand{\mysection}[1]{\stepcounter{mysection}\setcounter{equation}{0}
\setcounter{mysubsection}{0}\par\bigskip\noindent{\large\bf
\themysection .\ #1}\nopagebreak[4]\par\vskip .3cm}

\newcommand{\mysubsection}[1]{\stepcounter{mysubsection}
\par\medskip\noindent{\large\it
\themysection .\themysubsection\ #1}\nopagebreak[4]\par\vskip .3cm}
\def\l:{\mathopen{:}\,}
\def\r:{\,\mathclose{:}}

\def\inbar{\,\vrule height1.5ex width.4pt depth0pt}
\font\cmss=cmss12 \font\cmsss=cmss12 at 7pt
\def\IZ{\relax\ifmmode\mathchoice
{\hbox{\cmss Z\kern-.4em Z}}{\hbox{\cmss Z\kern-.4em Z}}
{\lower.9pt\hbox{\cmsss Z\kern-.4em Z}}
{\lower1.2pt\hbox{\cmsss Z\kern-.4em Z}}\else{\cmss Z\kern-.4em
Z}\fi}
\def\IB{\relax{\rm I\kern-.18em B}}
\def\IC{{\relax\hbox{$\inbar\kern-.3em{\rm C}$}}}
\def\ID{\relax{\rm I\kern-.18em D}}
\def\IE{\relax{\rm I\kern-.18em E}}
\def\IF{\relax{\rm I\kern-.18em F}}
\def\IG{\relax\hbox{$\inbar\kern-.3em{\rm G}$}}
\def\IP{\relax{\rm I\kern-.18em P}}

\def\IT{\relax{\rm T}}
\begin{document}

\begin{titlepage}

{\hbox to\hsize{hep-th/9806080 }}
{\hbox to\hsize{June 1998 \hfill UCSD-PTH-98-16}}
\bigskip

\begin{center}

\vspace{.5cm}

\bigskip

\bigskip

\bigskip

{\Large \bf   Branes with GUTs and Supersymmetry Breaking}

\bigskip

\bigskip

\bigskip

{\bf Joseph Lykken}$^{\bf a}$, 
{\bf Erich Poppitz}$^{\bf b,}$\footnote{AfterJanuary 1, 1999: Department
of Physics, Yale University, New Haven, CT 06520-8120, USA.}, 
and {\bf Sandip P. Trivedi}$^{\bf a}$ \\

\smallskip

{\tt lykken@fnal.gov, epoppitz@ucsd.edu, trivedi@fnal.gov}

\bigskip

\bigskip

$^{\bf a}${ \small \it Fermi National Accelerator Laboratory\\
  P.O.Box 500\\
 Batavia, IL 60510, USA\\}

\bigskip

$^{\bf b}${\small \it Department of Physics \\
University of California at San Diego\\
9500 Gilman Drive\\
La Jolla, CA 92093, USA}
 
\bigskip
 
\bigskip

{\bf Abstract}
\end{center}
{
We study Type I string theory compactified on a $\IT^6/\IZ_3$ orientifold. 
The low-energy dynamics is most conveniently analyzed
in terms of D3-branes. We show that a sector of the theory, which 
corresponds to placing an odd number of D3-branes at orientifold fixed points,
can give rise to an $SU(5)$ gauge theory with three generations of chiral 
matter fields.  The resulting model is not fully realistic, but the relative ease with 
which an adequate gauge group and matter content can be obtained is promising. 
The model is also of interest from the point of view of supersymmetry breaking. 
We show that, for fixed values of the closed string modes, the model breaks 
supersymmetry due to a conflict between a non-perturbatively generated 
superpotential and an anomalous $U(1)$ D-term potential. 
}

\end{titlepage}

\renewcommand{\thepage}{\arabic{page}}

\setcounter{page}{1}

\mysection{Introduction and Summary}

The past few years have seen remarkable progress in our understanding of 
the non-perturbative behavior of string theory \cite{sen}. D-branes have 
played a vital role in these developments \cite{tasibranes}. 
The consequences of this theoretical insight in 
string phenomenology  are just beginning to be explored. 
In this paper we attempt to take a few preliminary steps in this  direction. 
For related recent work see \cite{wittenstrongcy}--\cite{kaku}.
Perhaps the simplest idea to explore is that we live on a three dimensional
brane or somewhat more precisely, that the $3+1$ dimensional spacetime 
corresponds to the world volume of a set of D3 branes. This immediately 
gives rise to a question: can a (grand unified) theory accommodating the 
standard 
model interactions and matter content be obtained in this manner? 
 
D-brane model building is of interest from another point of view as well. 
Most of the model building so far has been carried out in the 
$E_8 \times E_8 $ heterotic string \cite{dienes}. 
In this context, there is a well known problem in reconciling, 
within the context of weakly coupled string theory,
the ``observed" 
unification of gauge coupling constants in supersymmetric 
extensions of the standard model \cite{mssmgut}  and the  value of Newton's 
constant. Witten \cite{wittenstrongcy} 
has recently suggested working with the strongly coupled
heterotic theory to avoid this problem. 
Another possibility, also mentioned
in \cite{wittenstrongcy}, is to consider model building in the Type I theory. 

We begin this paper 
by considering, in Section 2,  the question of gauge and gravity
unification in the Type I string theory. We show 
that both  the gauge coupling unification and the 
value of  the Newton constant  
can be obtained within the context of Type I  perturbation theory.
Moreover, the analysis indicates that in several cases the more 
appropriate  description is a T-dual one with D3 branes. 
This provides additional motivation
to enquire about the standard model arising from D3 branes.

In Section 3, we turn to this issue by considering a compactification
of the Type I theory on a $\IT^6/\IZ_3$ orientifold. This compactification 
has been considered earlier by \cite{sagnotti96}. 
We point out that, in addition to the sector
considered in \cite{sagnotti96}, the moduli space for this compactification 
has additional  disconnected branches, similar to the ones found in
\cite{wittentori}. The different branches correspond to distinct ways 
in which the branes can be placed at the various orientifold fixed points.
The additional branches of moduli space exhibit patterns of 
gauge symmetry  breaking that are not otherwise  allowed.
In particular, we show in Section 3.1, that  
an $SU(5)$ grand unified theory with three generations 
of matter fields in the ${\bf 10}$ and $\bar {\bf 5}$ representations 
can arise in this manner.  
In Section 3.2, 
we show that some 
nonperturbative consistency conditions \cite{Star}, \cite{wittentori}
leading to the existence of these additional branches are met.
The model obtained in 
this manner is not fully realistic: there are no Higgs fields present 
and  there are Yukawa couplings violating  
baryon and lepton number. 
Even so,  we view the relative ease with which an adequate gauge group and 
matter content can be obtained as encouraging.  

Finally, in Section 4, we turn to another aspect of the $SU(5)$ theory 
mentioned above. 
The theory has an additional $U(1)$ gauge symmetry,
which is anomalous. We show that a conflict between the 
non-perturbatively generated superpotential and the 
D-term of the anomalous $U(1)$ gives rise to 
supersymmetry breaking in this theory. Classically, the D-branes giving 
rise to the gauge theory are stuck at the orientifold plane. In the 
supersymmetry breaking vacuum, some of these branes are repelled by the 
orientifold and come to rest away from it.  In this discussion of 
supersymmetry breaking, we neglect the gravitational interactions and keep 
the dilaton and a relevant orientifold blow-up mode fixed. 
We show that  supersymmetry breaking occurs for any fixed values of the
dilaton and orientifold blow-up mode. Once these modes are taken to be 
dynamical, there are, as usual,  runaway directions along which supersymmetry 
is restored. What happens when the relaxation of the closed 
string modes and the gravitational interactions is included is an interesting 
question which we leave  for the future.

\mysection{Gauge Coupling  Unification on D3 Branes}

In this section, we discuss the constraints imposed on string model 
building by the requirement of gauge coupling unification 
(taking the values $\alpha_{GUT}$ and $M_{GUT}$
for supersymmetric extensions of the standard model) and the 
observed value of Newton's constant.
For the $E_8 \times E_8$ heterotic string, these requirements 
lead to the conclusion that string theory must be strongly 
coupled \cite{wittenstrongcy}.
In contrast, as has been noted earlier in \cite{wittenstrongcy}, we 
will see that in the case of the Type I string theory these 
requirements can be met while still
working at weak string coupling. 
Moreover,  the discussion below suggests that 
in several Type I models  
the six   compactified dimensions can have a  length somewhat bigger
than the inverse GUT scale. In these cases, the gauge group and charged
matter   would arise  from fields  living on D3-branes that fill the $3+1$
dimensional  flat spacetime.

The relation between the string scale $\alpha^{\prime}$,  Type I string 
coupling $g_I$,
volume of compactification  $V_6$, gauge coupling at unification
$\alpha_{GUT}= g^2 / 4 \pi$, and Newton's constant $G_N$
is given by \cite{sen}: 
\beq
\label{Newton}
G_N~=~{(2 \pi)^7 \over 16 \pi}~{\alpha^{\prime 4} \over V_6} ~g_I^2~,
\eeq
and
\beq
\label{alphagut}
\alpha_{GUT}~=~{(2 \pi)^7 \over 4 \pi}~ {\alpha^{\prime 3} \over V_6} ~g_I~ .
\eeq
Here we will consider the situation where the six compactified dimensions
have approximately the same size $R$. 
The volume $V_6$ is then roughly given by 
\beq
\label{defv}
V_6~=~( 2  \pi  R)^6
\eeq
To proceed,
we need to decide how to relate the unification scale 
$M_{GUT} \simeq 10^{16}$ GeV to $\alpha^{\prime}$ and $R$. 
In several string models the  gauge
couplings unify even in the absence of a grand unified group.
 In these cases, one
expects the  grand unification scale to correspond to the masses 
of the lightest extra
charged states present in the string theory.   
These extra states can be of two
kinds: Kaluza-Klein modes with a mass of order $1/R$, or higher string
modes with a mass of order  $1/\sqrt{\alpha^{\prime}}$. 
If we assume that $R > \sqrt{\alpha^{\prime}}$, the lightest extra states
have a mass $m \sim 1 / R$, leading to the 
relation $R \sim 1/ M_{GUT}$.
{}From eqs.~(\ref{Newton}), (\ref{alphagut}), and  (\ref{defv}) it then follows that: 
\beq
\label{wronga}
{\alpha^{\prime} \over R^2} ~=~ {1 \over \sqrt 2} ~\alpha_{GUT}~ R~ M_{Pl},
\eeq
and 
\beq
\label{gI}
g_I~=~4 \sqrt{2} ~{1 \over \alpha_{GUT}^2 ~R^3~ M_{Pl}^3 }.
\eeq
With the values $\alpha_{GUT} = 0.04$ and 
$R \sim 1/ M_{GUT} = (10^{16} {\rm GeV})^{-1}$ 
for the supersymmetric standard model \cite{mssmgut},
and 
$M_{Pl} = G_N^{-1/2} = 
1.2 \times 10^{19}$ GeV, we get from eq.~(\ref{gI}) that :
\beq 
\label{valgI}
g_I ~ \sim ~ 10^{-6}~ .
\eeq
Thus, the gauge coupling is small, as  mentioned above.  
However from  eq.~(\ref{wronga})  we find  that:
\beq
\label{watwo}
{\alpha^{\prime} \over R^2} \sim 34~ . 
\eeq
This  shows that our starting  assumption ($R > \sqrt{\alpha^\prime}$) 
about the  lightest  extra charged states coming 
from Kaluza-Klein modes is incorrect.  
A consistent solution is obtained by assuming that  
$R < \sqrt{\alpha^\prime}$.  
The  lightest  extra  states  which
enter at  the GUT scale are  then 
higher string modes with  mass $M \sim  1/\sqrt{\alpha^{\prime}}$.   
In this case,  the more  appropriate geometrical picture is 
obtained by  T-dualizing along the six compactified directions.   
Doing so turns the D9 branes into
D3 branes. The T-dual radius and string coupling are given by: 
\beq
\label{dualr}
{\tilde R} ~= ~{\alpha^{\prime} \over R} 
\eeq
and
\beq
\label{dualg}
{\tilde g_I}~ =~ \left( {\alpha^{\prime} \over R^2} \right)^3 ~g_I~.
\eeq
Eq.~(\ref{alphagut}) then  implies  directly that 
\beq
\label{valdualg}
{\tilde g_I} ~=~ 2 ~\alpha_{GUT}  \simeq  0.08~.
\eeq
Furthermore, since the lightest excitations are higher string modes we now 
set $\alpha^{\prime} \sim (M_{GUT})^{-2}$. 
Eq.~(\ref{Newton})  then  gives
\beq
\label{finalR}
{\tilde R} ~=~ {\sqrt{\alpha^\prime}}~ \left( {1 \over 8} ~
M_{Pl}^2 ~ \alpha^{\prime} ~{\tilde g_I}^2 \right)^{1 \over 6} 
\sim 3  M_{GUT}^{-1}~.
\eeq

It is useful to describe the resulting picture in words. The gauge group
and charged matter arises from D3-branes. 
The six compactified dimensions
have a length scale  somewhat bigger than the 
inverse GUT scale.  In particular, we note that
since all the degrees of freedom 
charged under the  gauge groups arise from open
strings that end on the three-branes, 
there are no momentum modes with mass of
order $1/{\tilde R}$ charged under the gauge group. 
Instead, there are
winding  modes with a mass of order ${\tilde R} /\alpha^{\prime}$ but these
are  somewhat heavier than the higher string modes with mass 
$\sim 1/ \sqrt{\alpha^\prime}$. 

We should emphasize that the  above  picture  is meant to be suggestive.
Whether it applies or not will depend on the details 
of the compactification. 
 It was noted in \cite{wittenstrongcy} that  in the $E_8 \times E_8$
theory  the large gauge coupling implies an extra dimension  at a scale
somewhat below the GUT scale (for recent work and a list of references,
see \cite{nilles}).
Here it is interesting to note  that  the 
presence of extra large dimensions might 
be true  in the Type I case as well, and more generally,
  in attempts to build string models involving branes.  
We should also  note that the  conclusion 
with regards to the smallness of the string 
coupling is secure regardless of the 
exact relation between $R$ 
and $\sqrt{\alpha^{\prime}}$.\footnote{For example, setting 
$\sqrt{\alpha^{\prime}} = R$ in eq.~(\ref{alphagut}) still gives  (\ref{valdualg})
for the gauge coupling.}

We end this section with a few comments. 
First, strictly speaking, the discussion above applies 
to models where   gauge
coupling unification occurs in the absence of a grand unified group. 
One can ask what happens  if the  low-energy  field theory  is  a grand unified 
theory.  In this case the  lightest extra string states need not occur at the 
GUT scale but could have  larger masses.  Eq.~(\ref{Newton}) then shows 
that  in these cases the compactification scale $\tilde R$  
should be comparable  to $\alpha^{\prime}$.  
Even so, as  we see in the next section it might be 
sometimes convenient to analyze  such a  model  in terms of  D3-branes. 
Second, in our analysis we have taken all the compact dimensions to have 
roughly the same size. This of course need not be true. For  recent 
discussions of large extra dimensions and weak scale strings see \cite{dim}.
Finally, we have assumed that the gauge group and matter content arises
from the perturbative sector of Type I theory. This, too, need not be true.
One could have a situation where some of the degrees of freedom arise from 
$9$-branes while others arise from $5$-branes; for 
model building along these lines see \cite{AFIZ}, \cite{kaku}, \cite{kakushadze}.

\mysection{A Three Generation Model on D3 Branes}

In this section, we present a simple ``three generation" model with D3 branes
placed at a $\IT^6/\IZ_3$ orientifold. 
The model is, admittedly, not a realistic one, but it will serve 
the purpose of making
several generic points quite explicit. 
In general, the moduli space of the gauge theory which governs the low-energy
 dynamics can be quite 
complicated with several disconnected sectors. The D3-brane picture allows for 
a geometric description of these different branches of moduli space 
\cite{wittentori}.
The different branches correspond to the distinct ways in which the branes
can be placed at the various orientifold fixed points. 

The additional branches of  moduli space 
can have multiple uses. We will see below that they exhibit interesting
 patterns of gauge symmetry breaking that are not otherwise  possible.
 In addition, branes placed at  
different  orbifold fixed points can serve as 
``visible" and ``hidden" sectors; the latter can
be responsible for supersymmetry breaking. The lightest excitations of
strings stretching between branes at different fixed points transform as 
fundamentals under both the ``hidden" and ``visible" gauge groups; these 
could be instrumental in communicating supersymmetry breaking. 

\smallskip

\mysubsection{The $\IT^6/\IZ_3$ orientifold}

We now turn to studying the Type I theory  compactified on  a $\IT^6/\IZ_3$
orientifold.  This theory has been analyzed  by \cite{sagnotti96} 
and more recently 
by \cite{kakushadze}, where the low energy dynamics 
was shown to correspond to an 
$SU(12) \times SO(8)$  gauge theory. 
Our main purpose here will be to study  some
sectors  of moduli space which are disconnected from the  $SU(12) \times
SO(8)$ theory mentioned above.  For this purpose it will be often convenient
to  T-dualize the Type I theory along the six directions of $\IT^6$. Doing so turns 
the 9-branes into  3-branes. The disconnected sectors  then correspond 
to placing  
an odd number  of 3-branes at  the orientifold  fixed planes and can be easily
visualized.   We show below how  an $SU(5)$ theory with three generations 
of matter fields in the  ${\bf 10}$ and $\bar {\bf 5}$ 
representations can be obtained in this 
manner. 

Let us describe the  $\IT^6/\IZ_3$ orientifold  in more detail.  We work for the most
part  in the   T-dual description involving  D3 branes.  The D3 branes 
stretch along
$X^{1,2,3}$. 
We introduce complex coordinates,   $z_1 = X^4 + i X^5, z_2 = X^6 + i X^7, 
z_3 = X^8 + i X^9$,  in the  compactified  six-dimensional space.
Consider the two-torus obtained  by identifying points under 
\beq
\label{twotorus}
z~ \simeq  ~z ~+ ~ R
  ~\simeq ~z ~+~ R ~ e^{i 2  \pi \over 3}~. 
\eeq
The $\IT^6$ is obtained by taking three copies of  this two torus, corresponding
to the three complex  coordinates  $z_1, z_2, z_3$. 
The orientifold group is  given by:
\beq
\label{orgroup}
G~=~\{~1, ~\alpha, ~\alpha^2 ,~ \Omega  R  (-1)^{F_L} ,
~ \Omega R (-1)^{F_L} \alpha,~
 \Omega R (-1)^{F_L} \alpha^2 ~\}~.
\eeq 
Here,  $\alpha$ is a spacetime symmetry  whose action is given by:
\beq
\label{orbifold}
( z_1, ~ z_2, ~z_3 ) ~\rightarrow ( \alpha  z_1, ~\alpha  z_2, ~\alpha  z_3 )~.
\eeq
$\Omega$ denotes world-sheet orientation reversal, and $R$ is a reflection 
$z_i \rightarrow - z_i, i = 1, 2, 3$. 
$F_L$ is an operator that flips the sign of the left-moving Ramond states. 
The orientifold group G  has a $\IZ_2$ subgroup
\beq
\label{gz2}
G_{\IZ_2}~=~ \{1 , \Omega  R  (-1)^{F_L} \}
\eeq
and a $\IZ_3$ subgroup
\beq
\label{gz3}
G_{\IZ_3}~=~\{1, \alpha, \alpha^2 \}~.
\eeq
These will play a useful role in the subsequent discussion. 

In addition to acting on the spacetime indices, 
the orientifold group acts on the
Chan-Paton indices of the open string states stretching between 
 D3 branes \cite{tasibranes}.
The action of the group elements,
$\Omega R (-1)^{F_L}$ and $\alpha$,  on the  Chan-Paton
factors $\lambda$ is:  
\beq
\label{eomega}
\lambda \rightarrow \gamma_{\Omega R (-1)^{F_L}} ~ \lambda^T ~
\gamma_{\Omega R (-1)^{F_L}}^{-1}~,
\eeq
and 
\beq
\label{egamma} \lambda \rightarrow \gamma_\alpha ~ \lambda ~
\gamma_\alpha^{-1}.  
\eeq 
The matrices   $\gamma_\alpha $ and  $\gamma_{\Omega R (-1)^{F_L}}$ must  
furnish a
representation of the orientifold group. 
The matrices  $\gamma_{\Omega R (-1)^{F_L}}$,
representing the action of the $\IZ_2$ part of the orientifold group
should obey \cite{tasibranes}:
\beq
\label{gammaomega}
\gamma_{\Omega R (-1)^{F_L}}\ =
\left( \gamma_{\Omega R (-1)^{F_L}} \right)^{T}~.
\eeq 
In the absence of the $\IZ_3$ orbifold projection,
 the $\Omega  R (-1)^{F_L}$ projection  would lead to 
 an $SO$ gauge group on the D3 brane world volume.

Tadpole cancellation conditions  play an important role in ensuring
the consistency of  the string compactification.  For the $\IT^6/\IZ_3$ orientifold
these were discussed in \cite{sagnotti96}.  
For the sake of brevity we will not discuss
a detailed derivation of these conditions here.  Instead we will content ourselves
with stating them; as the reader will see  these  conditions 
 give rise to anomaly free gauge theories. 

As expected,   the untwisted   Ramond-Ramond 4-form 
charge conservation conditions require 
the presence of 32 D3 branes to cancel the orientifold charge.  
In addition, there
are   charge cancellation conditions for the
twisted RR fields. 
Before stating these, it is useful to consider the  action of the
$G_{\IZ_3}$ and $G_{\IZ_2}$ subgroups of the orientifold group on the 
$\IT^6$.   Consider first a two torus shown in Fig.~1. 

\begin{figure}[ht]
\vspace*{13pt}
\centerline{\psfig{file=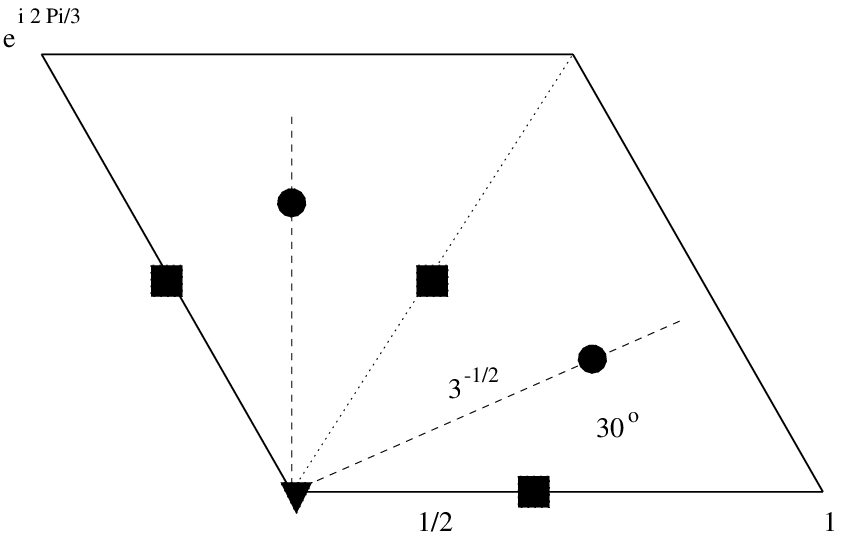}}
\vspace*{13pt}
{\centerline{\small Figure 1}}
\end{figure}

The  origin, denoted by the triangle in Fig.~1, is  the only fixed point 
with  respect to the full $\IZ_6$ orientifold group.  
In addition the $G_{\IZ_3}$
subgroup has two additional fixed points, at  
$z = {1\over \sqrt{3}} e^{i \pi\over 6}$
 and $z = {1\over \sqrt{3}} e^{i \pi\over 2}$, respectively. These two
fixed points are interchanged under the 
action of the $G_{\IZ_2}$ subgroup and are  denoted by
circles in Fig.~1. 
Similarly, the $G_{\IZ_2}$ subgroup  has three additional fixed points
at $z = {1\over 2}, {1\over 2} e^{i 2\pi\over 6}, {1\over 2} e^{i 2\pi\over 3}$, 
which are denoted by squares. 
These  are images of each other under the  $G_{\IZ_3}$ symmetry.  
The
fixed points for the $\IT^6$ can  now be deduced in  a  straightforward manner. 
There
is only one fixed point under the full $\IZ_6$ symmetry---the origin. 
 In addition,  as in the
$\IT^2$ case there are fixed 
points of the $G_{\IZ_3}$ symmetry which are transformed
into one another by  the $G_{\IZ_2}$ and vice-versa; these 
will be referred to, in what follows, as 
$G_{\IZ_3}$ and $G_{\IZ_2}$ fixed points, respectively. 

We can now return to the tadpole conditions for the 
twisted Ramond-Ramond fields.  If $\gamma_\alpha$ is the matrix 
which  represents the action of $G_{\IZ_3}$ on the branes at the origin, one
finds \cite{sagnotti96} that:
 \beq
\label{gammaalpha}
{\rm Tr} ~ \gamma_{\alpha} ~=~ -4 ~.
\eeq
In contrast, at a $G_{\IZ_3}$ fixed point one finds that 
\beq
\label{gammaz3}
{\rm Tr} ~ \gamma_{\alpha} ~=~ 0. ~
\eeq
Finally, at each $G_{\IZ_2}$ fixed point  (and consistently at its $\IZ_3$ image
points) one can choose to  place  an even or odd number of branes. 

The simplest way to meet these conditions is to place all  the $32$ D3 branes
at the origin. This gives rise, from eq.~(\ref{gammaalpha}),
 to a gauge theory
with $SU(12) \times SO(8)$
 gauge group with three generations of 
$(\overline{\Yfund}, \Yfund) + (\Yasymm, {\bf 1})$
 fields
which was discussed in \cite{sagnotti96}. 

The rank of the $SU(12) \times SO(8)$  gauge symmetry can be reduced by 
moving some of the branes away from the origin in a continuous manner.  To be
consistent with the  $\IZ_6$ orientifold symmetry,
however,  these branes can only be
moved  away from the  origin in sets of six. From  eq.~(\ref{gammaalpha}) it then
follows that the  rank of the $SU(N)$ factor  must 
always be odd; this  precludes an $SU(5)$  
gauge symmetry which is attractive from  a phenomenological point of view. 

\smallskip

\mysubsection{The  $SU(5)$ theory}

We turn now  to exploring some branches of moduli space,
which are disconnected 
from  the  $SU(12) \times SO(8)$ theory mentioned above.   
We will see  how  some of these branches give rise
to an $SU(5)$ gauge theory with three generations of  fields in 
the ${\bf 10}$ and $\bar{\bf 5}$ 
representations.   We discuss examples of such  disconnected  branches below,
but   before doing so it is worth summarizing the  essential
features responsible for the grand unified theory. 

In some sectors of moduli
space, an odd number of branes can be removed from the origin. 
In particular, a situation can arise where only  $11$  of the $32$ D3 branes  are 
left at the origin.    Eq.~(\ref{gammaalpha})  
then  implies that  one has   an  $SU(5)$ gauge symmetry. 
 In addition there  is an anomalous $U(1)$ symmetry  (which is 
broken at the string scale).  
The theory  has $N=1$ supersymmetry  with  matter 
content corresponding to three generations of matter fields which transform as: 
\begin{equation} \label{su5}
\begin{array}{c|c|c}
        &SU(5)&U(1)\\ \hline
A_{i = 1,2,3} & \Yasymm & 2 \\ \hline
 \bar{Q}_{i = 1,2,3} & \overline{\Yfund} & - 1\\
\end{array} ~.
\end{equation}
The theory has a renormalizable tree-level superpotential given by
\beq
\label{superpots}
W_{tree} = \epsilon^{ijk} A_i \bar{Q}_j \bar{Q}_k.
\eeq
The three generations arise  because the  $\IZ_3$ 
action in eq.~(\ref{orbifold}) does not distinguish between the three 
(complex) transverse coordinates,  thus one set of matter fields 
in eq.~(\ref{su5}) arise from each of them. 

We now turn to discussing how these disconnected branches arise. 
As we saw in the discussion above,  
if we start with all $32$ branes at the origin
and move  some away in a continuous manner, 
one is always left with an even
number  of branes at the origin.  
Thus to   get $11$ branes some of them must 
be placed at  fixed points of the  $G_{\IZ_3}$  of $G_{\IZ_2}$ subgroups, 
eqs.~(\ref{gz2}), (\ref{gz3}).  Now eq.~(\ref{gammaz3})
implies that  the number of branes at
  a $G_{\IZ_3}$ fixed point must be  a  multiple of three. 
 In addition, as we saw above, each  $G_{\IZ_3}$ fixed point has a $G_{\IZ_2}$
image. 
Thus,  one finds that  all the branes at a  $G_{\IZ_3}$ fixed point can be moved 
continuously  away in a $\IZ_6$ symmetric manner back to the origin. 
 Disconnected
branches of moduli  space  can be  obtained, however,
 by placing an odd number  of branes at a $G_{\IZ_2}$
 fixed point (and its two images under $G_{\IZ_3}$).   Since
there are a large  number of  $G_{\IZ_2}$ fixed points 
 in $\IT^6$ this gives rise to  a
large  number of possibilities. 
 We will not analyze all of them in detail here.  Rather, as an illustrative 
example, we focus on  a case   that gives rise to  the  
$SU(5)$ ``grand unified theory" mentioned above.

For this purpose, the simplest possibility is to consider a situation 
where all the branes 
are  at the origin  as far as the third  $\IT^2$ (corresponding to the 
$z_3$ coordinate) is concerned, but not as far as the 
other two tori are concerned.  Consider 
placing   one D3 brane at  a $G_{\IZ_2}$ fixed point in the first $\IT^2$ and 
at  a $G_{\IZ_2}$ fixed point in the second $\IT^2$---this 
brane has two images under the $\IZ_3$ symmetry. 
Next, place   one D3 brane at the  origin  of the first two-torus,
but at a  $G_{\IZ_2}$ fixed point of the second $\IT^2$---this 
brane has two images as well. 
Finally,  place  a  D3 brane  at a $G_{\IZ_2}$ fixed point of the first torus,
but at the origin  of the second. 
This brane has  two images as well.  
Altogether,  counting images, this gives us 
 $9$ D3 branes---an odd number---which 
are stuck to  orientifold planes  away from the origin (note, 
however, that the number of D3 branes at the 
$G_{\IZ_2}$ fixed points in each of the three two-tori is even; this is 
required by the consistency conditions discussed 
 in the following section). 
The remaining $23$ branes at the origin 
 give rise to an  $SU(9) \times SO(5)$ gauge symmetry. 
The matter fields transform as  three generations of 
 $(\Yasymm, {\bf 1})$ and 
$(\overline{\Yfund}, \Yfund)$, under  the $SU(9) \times SO(5)$ gauge 
group.\footnote{
We are using a notation where the vector of  $SO(N)$ is denoted by $\Yfund$.} 

Finally,  one can  move $12$---two sets of six---of the remaining branes
away from the origin, leaving behind $11$ branes, obtaining thus 
the $SU(5)$ theory mentioned  above, eq.~(\ref{su5}).   
We also  should mention that for  a generic
position of  these $12$ branes,
the full gauge symmetry includes an additional $U(1) \times U(1)$ factor.  
This  ``hidden sector"  gauge group can be further  enhanced
if the branes are placed at  $G_{\IZ_3}$ fixed points or at the $G_{\IZ_2}$ 
orientifold planes. 
For example, placing  six of the $12$ branes at a  $G_{Z_3}$ fixed 
point (and the
remaining  six at the image point) gives rise to  an $N=1$ theory with 
$SU(2)^3$ symmetry and  three sets of chiral matter  transforming  as  
bifundamentals under pairs of the  $SU(2)$'s.    Dividing the $12$ 
branes between  a $G_{\IZ_2}$  orientifold plane and its two images,
on the other hand, can give rise to  a theory with $N=4$ supersymmetry and 
an $SO(4)$ or  $SO(5)$ gauge symmetry  (the $SO(5)$ symmetry can
arise if the orientifold planes chosen already contain a D3 brane stuck
to them, as mentioned above in the discussion of the  disconnected
moduli space).  

So far, we have ignored the effects of open strings stretched between
branes at different fixed points. The lightest excitations of such
strings are massive states which transform as fundamental-antifundamental
under the respective world volume gauge groups. In the example we gave above, 
there can be two world volume theories with $N=1$ supersymmetry.
If supersymmetry were dynamically broken in one of these theories,
supersymmetry breaking would be communicated to the other gauge
theory via the massive chiral multiplets just described (and, of course, by the
supergravity in the bulk). A more precise investigation of this
 would probably  involve  details of the 
supersymmetry breaking dynamics and the stabilization of the  dilaton
\cite{dine}; we leave this for future investigation. 

\smallskip

\mysubsection{Non-perturbative consistency conditions}

There is one subtlety concerning disconnected sectors of moduli space
that needs to be mentioned. Sometimes such sectors are not allowed, 
even when they pass all the perturbative consistency conditions, 
due to non-perturbative reasons.  
Similar issues were addressed in \cite{Star}, \cite{wittentori}.   
We will not be able to discuss this matter in full detail here, 
but  will mention some salient points. 
The basic idea behind the non-perturbative  
consistency conditions is as follows.  
 The Type I  theory  does not have any  perturbative states which  transform
in spinor  representations  of $SO(32)$.  However, such states 
are present in the dual heterotic $SO(32)$ theory and 
are nonperturbative in the Type I theory. 
Allowing for such spinor representations imposes additional 
consistency conditions---whose origin from the Type I viewpoint is
non-perturbative.  

We can  verify that the example discussed above, 
giving rise to  the $SU(5)$ model,
meets  various   non-perturbative conditions. However, we should caution 
the reader  that there might be  other conditions, besides the ones  we have
checked that might not be met.\footnote{By way of comparison,  we note that 
the conditions  we have tested are  the analogue of 
 those discussed in Section 2.3 
of \cite{Star}.  The authors also formulated a stronger set of conditions in 
Section  5 of ref.~\cite{Star}.  We have not  investigated the  
presence of such stronger consistency conditions in  the present case.} 
It is in fact useful to carry out this discussion
in the original description of  Type-I theory in terms of $9$ branes.   
The essential feature giving rise to the disconnected branch of 
moduli space in the example above 
was the fact that there were 9 D3 branes (counting images) 
which were ``stuck" at the orientifold plane. 
 In the  T-dual 9-brane language we are using now, the positions of 
branes correspond to expectation values  for particular Wilson lines.  
The question is whether the Wilson lines' expectation values   
are consistent with the existence of states that transform as
$SO(32)$ spinors---the holonomies around
any contractable loop should be trivial in the appropriate  spinor
 representation.

The example discussed in Section 3.2  is 
equivalent to turning on four Wilson lines 
along the noncontractible loops of two of the two-tori.
It turns out in this case that the holonomy around any  contractible path 
is trivial in all representations of $SO(32)$. One can show this by explicitly
writing down the Wilson lines that correspond to the brane configuration
with $9$ D3 branes removed from the origin, which was described in Section 3.2.
Moreover,  in this  example,  an  explicit periodic flat connection, 
which is not constant on the  $z_1, z_2$  four-torus, 
can be found. This can be done by a straightforward 
generalization of  the construction of  ref.~\cite{rosly} 
to the case of $\IT^4$.  Furthermore, as in \cite{Star}, one can 
show that if $W$ is the  Wilson line  relevant for the 
particular $\IZ_3$ fixed point  then  
$(W \gamma_{\alpha})^3 =1$ in spinor representations as well. 

Before moving  on, let us mention that the example of Section 3.1,  giving 
rise to $11$ branes at the origin,
is  just one of several possibilities  consistent with the various conditions. 
For example, one can easily 
work out brane configurations with an odd total
 number of branes removed from the 
origin that involve moving branes to the $G_{\IZ_2}$ fixed points in all three
two-tori. 

\smallskip

\mysubsection{The GUT: shortcomings}

It is useful to describe the  construction of  the 
$SU(5)$ theory in  group theoretic terms perhaps 
more familiar to  some model builders.  
The $SO(32)$ gauge symmetry is broken to  an 
$SO(11)$ subgroup (times a hidden sector group).  The
orientifold projection then  further  breaks 
the symmetry to $SU(5)$  (with an additional  anomalous U(1)). 
 The $\bar {\bf 5}$ and ${\bf 10}$ matter fields arise 
from  the adjoint representation of $SO(11)$ by   the orientifold  projection. 
The three generations arise because there are
  three complex  (six real) transverse
dimensions and because the  orientifold group acts in an identical manner on the
three directions. 

The relative ease with which  a realistic gauge group and matter content can
be obtained in the Type I theory is interesting. We should note,
 that even though we used  a D3 brane description  
to simplify the  discussion, the construction 
as such was purely in the context of  perturbative Type I string theory. 
In particular, the  matter content was obtained,
  even  though we did not have any 
spinor representations of $SO(32)$ to begin with---in fact all the matter fields
can be thought of as being obtained  by 
truncating adjoint representations of  $SO(32)$. 

However, it should  also be noted  that the model is  meant as an illustrative
example  and  is not realistic.  
There are  several reasons for this. First,  there are no Higgs fields 
either  to break the $SU(5)$ gauge symmetry or to 
give rise to  the $SU(2)$ Higgs doublets of the 
supersymmetric  standard model.  
Second, and this is perhaps a more
important  limitation, 
as was mentioned in passing in eq.~(\ref{superpots}) above there 
is a Yukawa coupling in the theory that violates  baryon and lepton number.
The underlying reason for this coupling is that in the $N=4$ theory, which can 
be thought of as the starting point for the above construction, 
there is  a coupling 
involving the three adjoint fields corresponding to the transverse directions.  
In the case of, say, spinor representations of $SO(10)$, 
a trilinear ${\bf 16}^3$ coupling is not allowed by gauge invariance and 
an R-parity symmetry can often be imposed to prevent baryon- and 
lepton-number violating terms.  
However, in the present example, where all the matter arises
from adjoint representations no such R parity symmetry is present.  
This limitation is likely to be  quite  general.  

\mysection{Supersymmetry Breaking}

We turn now to another feature of the $SU(5)\times U(1)$ 
theory discussed above. 
  As we will see below, in the world-volume field theory context, a conflict 
 between  the  
non-perturbatively  generated  superpotential and 
the anomalous $U(1)$ D-term 
results  in   the  breaking of supersymmetry in this theory.  
  Our discussion of supersymmetry  breaking will  only involve  
the open string sector
corresponding to the  world volume theory on the D3 branes. 
 Gravity and other closed string  effects will be neglected.  In particular, 
 the dilaton and  the orientifold  blow-up mode \cite{dm}, which acts  
as the Fayet-Illiopoulos
term for the $U(1)$,  are regarded as coupling constants,
 and  will be  kept fixed in the discussion below.  
We will establish that supersymmetry breaking occurs
for  any finite value of these couplings.\footnote{ More accurately, 
supersymmetry breaking will
be shown when the string coupling is small enough to
argue with confidence that the low-energy dynamics is governed by 
the $SU(5) \times U(1)$ theory.}
But, as is usually the case, once  they are
allowed to vary, we find that  there are runaway directions along which 
supersymmetry is restored.  Perhaps, these could
 be stabilized by (yet poorly understood) nonperturbative 
corrections to the K\" ahler potential (see, e.g. the recent discussion in
\cite{dine} and references therein).
The stability of the supersymmetry breaking ground state in the context
of the full theory is a complicated issue, about which we have nothing to
say here.

Towards the end of this section, we will briefly 
comment on this runaway behavior and the gravitational back reaction. 
One final comment before we get started: the discussion 
in this section 
only involves the branes  at  the  $\IZ_6$ fixed point, 
 any details of compactification etc. are irrelevant  in this context. For
example,   the 
analysis here applies to the world volume theory of 11 D3 branes
on the noncompact $\IC^3/\IZ_3$ orientifold as well.

\smallskip

\mysubsection{Supersymmetry breaking in the $U(5)$ theory}

Our strategy to 
establish supersymmetry breaking is as follows.  We first neglect
the anomalous $U(1)$ and  show that in the $SU(5)$ theory  the resulting    
non-perturbative superpotential gives rise to runaway behavior.   Then on 
incorporating the anomalous $U(1)$ we find that its D-term gives rise 
to an energy that grows along the runaway
directions. This leads  to supersymmetry breaking. 

The  
$SU(5)$ ``three generation" model is an s-confining theory \cite{CSS}.
The infrared  degrees of freedom are the mesons and baryons
\beqa
\label{mesons}
C ~=~& A \cdot \bar{Q} \cdot \bar{Q} &~\sim~ ( {\bf 3},~ {\bf \bar{3}} ~,  0)~, 
\nonumber \\ 
B ~=~&A^5 &~\sim~ ({\bf 6}, ~{\bf 1} ~, 10)~, \\
 M ~=~&A^3 \cdot \bar{Q} &~ \sim~ ({\bf 8}, ~{\bf 3}  ~,  5)~, \nonumber 
\eeqa
where we have shown their transformation properties under the global
$SU(3)_A \times SU(3)_{\bar{Q}}$ symmetry and the last column in each entry
refers to the charges under the anomalous $U(1)$ which follow from  
eq.~(\ref{su5}).
The confining  superpotential is \cite{CSS}:
\beq
\label{wconfining}
W ~=~ { C_a^\alpha ~ B^{\beta \gamma} ~ M^{\delta a}_{\gamma} 
~\epsilon_{\alpha \beta \delta} ~+ ~ M^{\alpha a}_{\beta}~M^{\beta b}_\gamma
~M^{\gamma c}_\alpha ~\epsilon_{a b c} \over \Lambda^9} ~+ 
~\lambda ~ \delta^a_\alpha
~C_a^\alpha ~,
\eeq 
where $a, b, ... (\alpha, \beta, ...)$ 
denote indices under the $SU(3)_{{\bar{Q}} (A)}$ symmetry,
respectively, and the last term is the tree-level superpotential.
The tree-level
superpotential breaks the global symmetry to the diagonal $SU(3)_{diag}$.
It lifts  all the  $C$ flat directions, but  does not lift  the $B$ and  
some of the  $M$   directions.  The
superpotential coupling $\lambda$ in (\ref{wconfining}) is proportional to the 
value of the gauge  coupling at the string scale (since the tree-level 
superpotential is the projection of the $N=4$ superpotential).

We will show now that the F-term equations of motion following from 
(\ref{wconfining}) have no solutions for finite field expectation values. 
Consider the equations
of motion following from the superpotential (\ref{wconfining}) (suppressing 
numerical constants):
\beq
\label{eomC}
M_\alpha^{\beta a} ~B^{\alpha \lambda}~\epsilon_{\beta \lambda \delta}~=~ 
\delta^a_\delta ~,
\eeq
\beq
\label{eomM}
\epsilon_{a b c}~M^{\beta b}_\alpha ~M^{\gamma c}_\beta ~+~\epsilon_{\alpha
\beta \delta} ~C_a^\beta ~B^{\delta \gamma} ~= ~0~,
\eeq
\beq
\label{eomB}
\epsilon_{\gamma \alpha \beta}~C_a^\alpha ~M_\delta^{\beta a} ~= ~0~.
\eeq
Multiplying the first equation (\ref{eomC}) by 
$\epsilon^{\delta \mu \nu} M_\nu^{\gamma d} \epsilon_{bad}$, 
summing over $\delta$ and $a$, and substituting for
$\epsilon_{a b c} M^{\beta b}_\alpha M^{\gamma c}_\beta $ from (\ref{eomM}),
we obtain:
\beq
\label{eq1}
-~\epsilon_{\alpha \delta \beta}
~B^{\alpha \mu}~ B^{\delta \gamma} ~C_b^\beta ~
+ ~ \epsilon_{bad} ~M^{\mu a} ~B^{\alpha \nu}~M_\nu^{\gamma d} ~ = ~
M_b^{\gamma \mu} ~-~\delta_b^\mu ~M_\nu^{\gamma \nu}~.
\eeq
Now under $SU(3)_{diag}$, the field $M_\beta^{\alpha a}$ decomposes as a
${\bf 3}$, given by the partial trace  $M^{\alpha a}_a$, a ${\bf \bar{6}}$ 
which is antisymmetric in the upper two indices,
and a ${\bf 15}$ which is  symmetric in the upper indices (and traceless). 
Note that the l.h.s. of (\ref{eq1}) is antisymmetric in $\mu, \gamma$,
hence only
the r.h.s. contributes to the symmetric part of $M$. This gives rise to 
the relation:
\beq
\label{symm}
M^{\gamma \mu}_b~+~M^{\mu \gamma}_b ~- 
~\delta^{\mu}_b~ M^{\gamma \nu}_{\nu} ~
-~\delta^{\gamma}_b~ M^{\mu \nu}_{\nu}~=~0~,
\eeq
from which in turn it follows that $M^{\alpha a}_a$ vanishes as does the 
symmetric part  of $M$. Thus the ${\bf 3}$ and ${\bf 15}$ 
components of $M$ are zero.  The remaining ${\bf \bar{6}}$ can be written as
\beq
\label{assym}
M^{\gamma \mu}_a~=~ \epsilon^{\gamma \mu \kappa} ~s_{\kappa a}~,
\eeq
where $s$ is symmetric in the two indices. 
Substituting into eq.~(\ref{eomC}), and evaluating for $B\cdot s$ then leads
to the  relation:
\beq
\label{invert}
s ~= ~- {1 \over 2} ~B^{-1}~.
\eeq 
Substituting eq.~(\ref{assym}) in (\ref{eomB})
 and noting that $s$ is invertible, eq.~(\ref{invert}) then leads to 
$C_a^\alpha = \delta_a^\alpha {\rm Tr} ~C$. This  then implies that $C = 0$.
Finally, substituting into eq.~(\ref{eomM}), one similarly finds that 
$s_{\alpha \beta}  = \delta_{\alpha \beta} ~{\rm Tr}~ s$. This leads to the 
conclusion that
$s = 0$. But now we see from eq.~(\ref{invert}) that $B$ must go to infinity.
Thus we have established that there are no 
solutions to the F flatness conditions at finite expectation values. 

The equations (\ref{eomC})--(\ref{eomB}) do have runaway solutions. 
The discussion above leads to the conclusion that along a runaway direction,
$C$ and $M \rightarrow 0$, while 
$B \rightarrow \infty$ in an invertible manner---more precisely
$B^{-1} \rightarrow 0$.
For example, a  runaway vacuum solution 
with $SU(3)_{diag} \rightarrow SO(3)$ global
symmetry  is $B^{\alpha \beta} \sim \delta^{\alpha \beta} b$,
$C^{\alpha a} \sim \delta^{\alpha a} b^{- 3}$,
$M^{\alpha a \beta} \sim \epsilon^{\alpha a
\beta} b^{-1}$, with $b \rightarrow \infty$.
 The physics along 
the $B$, det $B \ne 0$ directions is easy to understand. Along these directions
the mesons $C$ and $M$
obtain mass. Upon integrating them out, 
 the superpotential of the low-energy theory 
is $W_{eff} = \lambda^3 \Lambda^{18}/{\rm det} B$, 
showing explicitly the runaway behavior.

So far we have studied the non-perturbative superpotential. Now let us 
include the anomalous $U(1)$ by ``turning it on" in the effective theory 
of the mesons $C$, $M$, and $B$. 
The last column in each row of eq.~(\ref{mesons}) 
gives the $U(1)$ charges of the three fields. 
We see that $C$ has charge zero, while $B$
and $M$ both have positive charge. 
We have argued above that along a runaway
direction $B$ must go to infinity (in an invertible manner) while $M$ and $C$
go to zero.
Since all components of $B$ have   positive charge with respect to the $U(1)$,
we see that along such a direction the $U(1)$ D-term contribution to the energy
blows up. Thus, the runaway behavior dictated by the non-perturbative 
superpotential is in conflict with the $U(1)$ D-term potential, leading to the 
breaking of supersymmetry (note that 
this is similar to the mechanism of ref.~\cite{anomalousu1}).

\smallskip

\mysubsection{Remarks on supersymmetry breaking}

We end this discussion of supersymmetry breaking with a few remarks. 
We first 
remind the reader of an important feature of type I orbifolds: 
the anomaly cancellation
of the ``anomalous" $U(1)$s is provided, as in the heterotic case, by a
Green-Schwarz mechanism. However, unlike the heterotic case, 
the axion that shifts under the $U(1)$ to cancel the anomaly is a
model-dependent field---the twisted Ramond-Ramond field from the 
closed string sector (this has been pointed out in \cite{dm} and recently
discussed in \cite{IRU}).  It is in the same 
supermultiplet as the orbifold blow-up
mode (the twisted NS-NS field) and can be described in terms of a chiral
superfield, denoted hereafter by $C$, with a kinetic term
\begin{equation}
\label{RRlagr}
\int ~d^4 \theta ~\left( C ~ + ~ C^\dagger ~ + V \right)^2 ~ + ~ \ldots ~,
\end{equation}
where dots denote higher-order terms.
The leading 
term (\ref{RRlagr}) can be written by demanding
$U(1)$ invariance and a smooth  kinetic term for
 $C$ in the orbifold limit $\langle C \rangle = 0$. 
Here  $V$ is the anomalous $U(1)$ vector superfield; in addition to (\ref{RRlagr}),
the field $C$ also has a Wess-Zumino coupling to the gauge field strengths, of
the form $\int d^2 \theta C W^\alpha W_\alpha$ \cite{dm}. 
In a superunitary gauge, the term (\ref{RRlagr}) represents a mass term
(of order the string scale)  for the anomalous $U(1)$ vector superfield. By 
giving an expectation value to the real part of $C$ (blowing up the orbifold)
one can induce ``tree-level" FI terms, with 
$\zeta_{FI}^2 \sim  \langle C + C^\dagger \rangle$, as follows from (\ref{RRlagr}).
That (\ref{RRlagr}) is correct follows from the computation of  
ref.~\cite{dm} of  the coupling of the real part of $C$ (the twisted NS-NS field)
to the D-term of the vector superfield (and from a subsequent 
supersymmetry transformation). This coupling
arises from the disk with two scalar vertex operators attached to the boundary,
and a closed string twisted NS-NS scalar vertex operator in the bulk \cite{dm},
and is of order ${\cal{O}}(g_{string}) \sim g^2$. 

We note that the conclusion regarding supersymmetry breaking is 
true  for any sign and finite value of the $U(1)$ Fayet-Illiopoulous term. 
Depending on the sign of the FI term, the D-term potential can have a zero
at finite values of the fields.\footnote{Since the $U(1)$ is anomalous one
expects that the 
FI term is renormalized at one loop. However, an explicit calculation
shows that the FI term is not generated in open string orbifolds, because
of cancellations between contributions of worldsheets of different topology
\cite{FItype1}.} However the 
F-term potential vanishes only at infinity. Thus supersymmetry is 
broken. It is possible 
though  to have vanishing  D- and F-term potentials for 
infinite (negative) value of the FI term.  

One would like to find out where the resulting supersymmetry breaking 
vacuum lies. Unfortunately, this is quite difficult---as the following 
argument shows,
one expects the vacuum to lie in a strongly coupled region 
where a semiclassical analysis is not applicable. 
Assuming first that such an analysis is valid, upon
balancing the U(1) D-term energy with the 
F-term energy one finds that the vacuum energy (with vanishing FI term)
 scales as  $g_1^{16/9} \lambda^{2/3} \Lambda^4$, while the typical 
 expectation value of a field goes like 
$v \sim \lambda^{1/6} g_1^{-1/18} \Lambda$. 
Here $g_1$ is the $U(1)$ gauge coupling, 
$\lambda$ is the tree-level Yukawa coupling, eq.~(\ref{wconfining}),
and $\Lambda$ is the strong coupling scale of the $SU(5)$ gauge theory.
For a semiclassical analysis to be valid, one requires $v \gg \Lambda$.
 If the gauge coupling $g_1$ and $\lambda$ were independent parameters,
 this could have been achieved by taking $g_1$ to zero
 keeping $\lambda$ fixed. However, in our case 
$g_1 \sim \lambda$, thus for small $g_1$ the vacuum lies in the strongly 
coupled region and the semi-classical analysis
 is not applicable. One could make $v \gg \Lambda$ 
by taking $g_1 \gg 1$, but then the string coupling would be 
large, again making a semiclassical analysis invalid. 

While we cannot determine the vacuum explicitly, we know that 
some of the $B$, $M$, and $C$ fields must get vacuum expectation 
values. 
These expectation values should correspond to 
displacing some of the D3 branes
away from the orientifold. We remind the reader that the 
$SU(5)$ theory under consideration here is the 
world volume theory for $11$ D3 branes placed at the 
orientifold. Classically, the 11 D3 branes are all 
stuck to the orientifold
plane and the configuration has no moduli.  This corresponds to the fact 
that in order to 
meet the tadpole conditions and respect 
the $\IZ_6$ symmetry no branes
can be moved away from the orientifold point.  
Quantum mechanically, due to non-perturbative 
supersymmetry breaking effects we see that some of the branes are repelled
by the orientifold plane and come to rest away from
 it so as to minimize the 
energy. 
Since there are no moduli
the  configuration of D3 branes
cannot be described in terms of classical geometry.  
The displacement of branes which are 
classically stuck at the orientifold is somewhat reminiscent of the 
splitting of orientifold 7-planes discussed in ref.~\cite{SSX}. 

As was mentioned in the beginning of this section, the above analysis 
neglected
all interactions with closed string sector modes. One might at first expect 
that gravitational interactions are small at low-energies and so can be 
neglected. But once supersymmetry is 
broken and a (boundary) cosmological 
constant is induced, this is not apriori true. 
Also, interactions with some other closed 
string modes, which determine the 
couplings of the brane theory are important. 
There are two modes of this 
kind. A blow-up mode for the orientifold determines
 the FI term of 
the $U(1)$ \cite{dm}; this mode (together with its partner) also enters in the 
determination of the coupling constant (theta angle)  
for the $SU(5)$ theory. 
Similarly the dilaton determines the gauge coupling 
of the $U(1)$ and together with the blow-up mode 
mentioned above determines the $SU(5)$ gauge coupling. 
As we have argued here supersymmetry 
breaking occurs for any fixed values of 
these couplings, but there are runaway directions 
along which it can be restored. For example,
 as we saw above there is a direction along which the FI term can
go to infinity with appropriate sign. 
Similarly, if the dilaton goes to infinity, supersymmetry is restored.
In fact, as has been argued recently in \cite{dine}, it is necessary
to include the dynamics of these closed string sector modes to get a complete 
description of supersymmetry breaking.  Without this there is no goldstino,
which signals that the description of supersymmetry breaking is incomplete.   
 It is interesting to ask how the  system  of D3-branes will evolve once
the dilaton and the blow-up mode are allowed to 
relax and the gravitational 
back reaction is put in. We leave this question for the future. 

Finally, another natural  configuration to consider involves not $11$, but $8$ 
branes
placed at the $\IZ_6$ orientifold plane. This is the minimum number required to
meet the tadpole conditions (e.g. starting with $32$ branes and moving
$24=6 \times 4 $ away leaves us with $8$ branes). The corresponding 
theory has an $SU(4)\times U(1)_A$ gauge symmetry and three generations of 
fields which transform in the $\Yasymm $ representation of the group. 
There is in addition an anomalous $U(1)$ under which each of the $\Yasymm$ 
fields has the same charge. In this case if the  FI term (i.e. the 
orientifold blow-up parameter) vanishes,  supersymmetry is unbroken.  
This is because, in contrast to the $SU(5)$ model, the $SU(4)$ theory has
a branch of moduli space where no dynamical superpotential is 
generated---this can be inferred from
 \cite{IS} by noting that the $SU(4)$ theory 
with three ${\bf 6}$'s  is 
equivalent to the $SO(6)$ theory with three vectors. 
The breaking of 
supersymmetry is then purely due to the $U(1)$ D-term and vanishes for 
vanishing FI parameter.

\mysection{Acknowledgments}

We would like to thank Ken Intriligator and  Witek Skiba for discussions.
E.P. was supported by DOE contract no. DOE-FG03-97ER40506. The research
of JL and ST is supported by the Fermi National Accelerator Laboratory,
which is operated by the Universities research Association, Inc., under
contract no. DE-AC02-76CHO3000.

\nc{\ib}[3]{ {\em ibid. }{\bf #1} (19#2) #3}
\nc{\np}[3]{ {\em Nucl.\ Phys. }{\bf #1} (19#2) #3}
\nc{\pl}[3]{ {\em Phys.\ Lett. }{\bf B#1} (19#2) #3}
\nc{\pr}[3]{ {\em Phys.\ Rev. }{\bf D#1} (19#2) #3}
\nc{\prep}[3]{ {\em Phys.\ Rep. }{\bf #1} (19#2) #3}
\nc{\prl}[3]{ {\em Phys.\ Rev.\ Lett. }{\bf #1} (19#2) #3}

\end{document}